\documentclass[reviewcopy]{elsarticle}
\usepackage{natbib,longtable}
\usepackage{graphicx,array}
\usepackage[reviewcopy]{adndt}

\begin{document}

\begin{frontmatter}

  \title{Energies and E1, M2 transition rates for Mo XXX}

\author[xzit]{Feng Hu\corref{cor1}}
\ead{hufengscu@139.com,Tel.+86-0516-83201360}
\author[xzit]{Yan Sun}
\author[xzit]{Maofei Mei}

\author[caep]{Jiamin Yang}
\cortext[cor1]{Corresponding author}
\address[xzit]{School of Mathematic and Physical science, Xuzhou Institute of technology, Xuzhou, 221111 Jiangsu,People's Republic of China}

\address[caep]{Research Center of Laser Fusion, China Academy of Engineering Physics, Mianyang,  621900 Sichuan, People's Republic of China}

\begin{abstract}
Based on relativistic wavefunctions from multiconfigurational Dirac-Hartree-Fock (MCDHF) and configuration interaction calculations, 
energy levels, radiative rates, and wavelengths are evaluated for all levels of 3s$^2$3p, 3s3p$^2$, 3s$^2$3d, 3p$^3$, 3s3p3d, 3p$^2$3d and 3s3d$^2$
configurations of Al-like Molybdenum ion (Mo XXX). Transition probabilities
are reported for  E1 and M2 transitions from the ground level. The valence-valence and core-valence correlation effects are accounted for in a systematic way. Breit interactions and quantum electrodynamics effects are estimated in subsequent relativistic configuration interaction calculations.  Comparisons are made with the available data in the literature
and good agreement has been found which confirms the reliability of our results.
\end{abstract}


\end{frontmatter}
\section{Introduction}

The spectra of Al-like ions(Z$>$30) have received a great deal of attention both experimentally and theoretically. The Al-like molybdenum studied in this paper is no exception. Molybdenum has many applications in different scientific fields. For example, molybdenum can be used as component of plasma-facing material in the Alcator C-Mod reactor\cite{Malcheva2006} or the experimental advanced superconducting tokamak\cite{Liu2010}. These applications need a large amount of atomic parameters to describe the different ionization degrees of Mo. But for Al-like molybdenum, radiative data have only been published from few works.

In the experimental front, spectra of Mo XXX generated in a laser-produced plasma were observed from 10 to 190 {\AA} by Burkhalter \emph{et al}\cite{Burkhalter1977}. The spectra of Mo formed by laser beam irradiation of solid molybdenum targets were produced by Mansfield \emph{et al}\cite{Mansfield1978}. Transitions of the 3s$^2$3p$^k$-3s3p$^{k+1}$ and 3p$^k$-3p$^{k-1}$3d transitions of molybdenum were identified in the Princeton large torus tokamak by Finkenthal \emph{et al} \cite{Finkenthal1985}. Wavelengths of the transitions of 3s$^2$3p-3s3p$^2$ and 3s$^2$3p-3s$^2$3d in the Mo XXX were reported by Hinnove \emph{et al}\cite{Hinnov1986}.Al-like spectra of molybdenum generated in a tokamak plasma was recorded by Sugar \emph{et al}\cite{Sugar1988}. Detailed analysis of the n=3,$\Delta$n=0 transitions in Mo XXX from the JET tokamak plasmas were made by Jup\'{e}n \emph{et al}\cite{Jup1990}.

In the theoretical front, Farrag  \emph{et al} used the relativistic wave functions obtained by the parametric-potential method to study the trends of energy levels and oscillator strengths for electric-dipole for Al-like ions\cite{Farrag1981,Farrag1982}. Sugar and Kaufman performed three parameters method to calculate the wavelength and transition rates for magnetic-dipole transitions within 3s$^2$3p for Mo XXX by \cite{Sugar1984}. Huang applied the multiconfiguration Dirac-Fock(MCDF) technique to present energy levels and wave functions\cite{Huang1986}.  Gebarowski \emph{et al} gave a relativistic multiconfiguration Dirac-Fock study of 3s$^2$3p-3s$^2$3d transition in aluminum isoelectronc sequence\cite{Gebarowski1994}. Lavin \emph{et al} reported theoretical oscillator strengths for 3s$^2$3p $^2$P-3s$^2$3d $2$D, 3s$^2$3p $^2$P-3s$^2$4s $^2$S and 3s$^2$4s $^2$S-3s$^2$4p $^2$P using the quantum defect orbital method (QDO) and its relativistic counterpart (RQDO)\cite{Lavin1997}. Charro \emph{et al} analysed the trends in E2 and M1 transition rates between 3p$_{3/2}$ and 3p$_{1/2}$ levels in 3s$^2$3p$^k$ systems using RQDO method\cite{Charro2003}. Safronova \emph{et al} displayed a computation of relativistic many body calculations of electric-dipole properties for n=3 in Al-like ions\cite{Safronova2003}. Hao \emph{et al} investigated the energy levels in Al-like ions by using the Multiconfiguration Dirac-Hartree-Fock(MCDHF) method\cite{Hao2010}.

Tr\"{a}bert conducted a critical assessment of theoretical calculations of structure and transition probabilities from a experimenter's view\cite{Trabert2014}. He pointed out that new computations can match measurement, fill gaps and suggest revisions closely with almost spectroscopic accuracy. And these citations of theoretical work as well as the ones
for experimental data are certainly incomplete. For example, limited energy levels and transitions were considered\cite{Hao2010}, or limited transitions 3s$^2$3p-3s3p$^2$\cite{Hao2010} and 3s$^2$3p-3s$^2$3d\cite{Gebarowski1994}.  So in this paper, the large-scale multiconfiguration Dirac-Hartree-Fock(MCDHF) method is performed to calculate the E1,
and M2 wavelengths, oscillator strengths, transition
probabilities and fine-structure levels for Mo XXX using the new release\cite{Jonsson2013} of the GRASP2K code\cite{Jonsson2007}. Eight configurations (1s$^2$2s$^2$2p$^6$)3s$^2$3p, 3s3p$^2$, 3s$^2$3d, 3p$^3$, 3s3p3d, 3p$^2$3d, 3s3d$^2$ and 3p3d$^2$ are included in this calculation. On the basis of our previous work\cite{Hu2010,Hu2011}, in this paper, the valence-valence(VV) and core-valence(CV) correlation effects are taken into account in a systematic way.
Breit interactions and
quantum electrodynamics(QED) effects have been added. This computational approach enables us to present a
consistent and improved data set of all important transitions of the Mo XXX spectra, which
are useful for identifying transition lines in further investigations.
\section{Method}
\subsection{Theory}
The MCDHF method has recently been described in great detail by Grant\cite{Grant2007}. Hence we only repeat the essential features here. Starting from the Dirac-Coulomb Hamiltonian
\begin{equation}\label{eq:1}
  H_{DC}=\sum_{i=1}^N(c\alpha_i\cdot\textbf{p}_i+(\beta_i-1)c^2+V_i^N)+\sum_{i>j}^N\frac{1}{r_{ij}}
\end{equation}
where $V^N$ is the monopole part of the electron-nucleus Coulomb
interaction, the atomic state functions (ASFs) describing different
fine-structure states are obtained as linear combinations of symmetry
adapted configuration state functions (CSFs)
\begin{equation}\label{eq:2}
  |\gamma J M_J\rangle=\sum_{j=1}^{N_{CSFs}}c_j|\gamma_j J M_J\rangle
\end{equation}

In the expression above J and $M_J$ are the angular quantum numbers.
$\gamma$ denotes other appropriate labeling of the configuration
state function, for example parity, orbital occupancy, and coupling
scheme. The configuration state functions are built from products
of one-electron Dirac orbitals. In the relativistic self-consistent
field (RSCF) procedure both the radial parts of the Dirac orbitals
and the expansion coefficients are optimized to self-consistency.
The Breit interaction

\begin{equation}\label{3}
  H_{Breit}=-\sum_{i<j}^N[{\alpha_i}\cdot{\alpha_j}\frac{cos(\omega_{ij}r_{ij}/c)}{r_{ij}}+(\alpha_i\cdot\nabla_i)(\alpha_j\cdot\nabla_j)
  \frac{cos(\omega_{ij}r_{ij}/c)-1}{\omega_{ij}^2r_{ij}/c^2}]
\end{equation}
as well as leading quantum electrodynamic (QED) corrections can
be included in subsequent relativistic configuration interaction
(RCI) calculations\cite{McKenzie1980}. Calculations can be done for single levels,
but also for portions of a spectrum in the extended optimal level
(EOL) scheme, where optimization is on a weighted sum of energies\cite{Dyall1989}
. Using the latter scheme a balanced description of a number
of fine-structure states belonging to one or more configurations
can be obtained in a single calculation.

\subsection{Generation of configuration expansions}
Different correlations were included into the calculation in a systematic approach. The correlation energy is defined as the energy difference between the exact solution to the Dirac equation and the DF solution. The contribution from different types of correlation then can be defined as the energy difference between the solution including the particular correlation under investigation and the DF solution. To classify the correlation, the atomic electrons can be divided into two parts: valence electrons and core electrons\cite{Grant2007}. As a result, the correlation between the valence electrons is defined as valence correlation (VV), and the correlation between the valence electrons and core electrons is defined as core-valence correlation (CV)\cite{Grant2007}.

It is, from some perspectives, desirable to perform separate calculations for each of the studied atomic states. This approach, however, is impractical and time consuming. Instead the atomic state functions for a number of closely spaced levels were determined together in the so-called extended optimal level (EOL) procedure\cite{Dyall1989}. To account for the close degeneracy between 3s$^2$3p and 3s3p$^2$, the atomic state functions for 3s$^2$3p $^2$P$^o_{1/2}$ and $^2$P$^o_{3/2}$, 3s3p$^2$ $^4P_{1/2,3/2,5/2}$, $^2P_{1/2,3/2}$, $^2D_{3/2,5/2}$  and $^2S_{1/2}$, were determined simultaneously. In the remaining cases atomic state functions for levels belonging to the same configuration were grouped together.

In the MCDHF approach, the correlation is represented by different constraints on the generation of the CSFs included in equation~\ref{eq:2}. If we only include the VV, the core electrons are kept fixed in all the CSFs generated. To include CV, we allow one of the core electrons to be excited to generate the CSFs.

\subsection{Calculation procedure}
As a starting point MCDHF calculations in the EOL scheme were performed for each group of atomic states using configuration expansions including all lower states of the same \emph{J} symmetry and parity, and a Dirac-Coulomb version was used, for the optimization of the orbitals, including Breit corrections in a final configuration interaction calculation\cite{Grant2007}. To build a CSF expansion, the restrictive active space methods were also used. The idea of the active space methods is to consider only electrons from the active space and to excite them from the occupied orbitals to unoccupied ones. The orbital was increased systematically in order to monitor the convergence of the calculation. Since the orbitals with the same principal quantum number n often have similar energies, the active set is usually enlarged in steps of orbital layers. It is convenient to refer to the \{1s, 2s, 2p\} set of orbitals as the n = 2 orbital layer, \{1s, 2s, 2p, 3s, 3p, 3d\} as the n = 3 layer, etc. Larger orbital sets can result in a considerable increase of computational time required for the problem, and appropriate restrictions may be necessary. We divided up the calculations into two parts, one where we optimized a set of orbitals for the even states and one for the odd states, i.e. the upper and lower states were described by two independently optimized sets of orbitals. Because of this we had to use biorthogonal transformation~\cite{Olsen1995} of the atomic state functions to calculate the transition parameters.

In our calculations, we generate the CSFs using the active space approach, we do this by exciting electrons from the spectroscopic reference configuration to a set of orbitals called the active set (AS). The active set is a set of orbitals which are all orbitals except those common to all CSFs, and it defines the CSFs included in the ASF. We increase the AS in a systematic way to ensure the convergence of the atomic parameters under consideration.

Some tests were undertaken for Mo XXX, to determine what sort of corrections is necessary to be included in our calculation. First, we only included the VV . In subsequent calculations, the CV correction due to the 1s, 2s, and 2p orbitals was successively included. The results of these tests show CV makes significant changes to the calculations and cannot be ignored. Furthermore, the largest contribution is due to the CV correction from the 2p orbitals, and the correction from the 1s and 2s CV correction is very small. Thus, like our previous papers, we only include the CV of 2p in calculation.

The similar calculation procedure have been introduced in ref\cite{Hu2011}.  For Al-like ions, the ground and first excited configurations are 3s$^2$3p, and 3s3p$^2$ respectively. In the first step, the active set(AS) is

\begin {equation}
AS1=\{3s,3p,3d\}
\end {equation}

Then, the active set was increased in the way shown as follows:
\begin {equation}
AS2=AS1+\{4s,4p,4d,4f\}
\end {equation}
\begin {equation}
AS3=AS2+\{5s,5p,5d,5f,5g\}
\end {equation}
\begin {equation}
AS4=AS3+\{6s,6p,6d,6f,6g\}
\end {equation}

The VV and CV used different active set, here we discussed each clearly.

Here, in our VV method, we set $1s^22s^22p^6$ as our core electrons in the calculation.
Then we considered to increase the principal quantum number n, and optimized the orbitals AS1, AS2, AS3, and AS4.

In CV model, the core electrons is $1s^22s^22p^5$, then we optimized the layer by n. We generate the CSFs  of
the form of  $1s^22s^22p^5ASn$, n=1-4. Also, the CSFs of CV have the form of $1s^22s^12p^6ASn$, n=1-4.

\section{Results and discussion}
The success of a calculation relies on a judiciously chosen
configuration expansion\cite{Sturessona2007}. To ensure the convergence of
a calculated expectation value within a certain correlation
model, the configuration expansion must be enlarged in a
systematic way. A very efficient way of doing this is to
use the active set approach, where jj -coupled configuration
state functions of a specified parity P and angular momentum. J symmetry are generated by excitations from one or more
reference configurations to an active set of orbitals. The
convergence of the atomic property can then be studied
as a function of the size of the active set. The GRASP2K procedure JJ2LSJ\cite{Jonsson2013} was used for the transformation of ASFs (atomic
state functions) from a jj-coupled CSF basis into a LSJ-coupled CSF basis\cite{Naze2013} for the results.
 The calculated energy levels (in cm$^{-1}$) are shown in descending order in table \ref{tab1} where also comparisons with experimental results obtained from NIST levels\cite{NIST}  and theoretical  values from Hao \cite{Hao2010} are included.  Our calculated level energies for Mo XXX agree well with the NIST levels.
 As can be seen from table \ref{tab1}, the VV correlations have converged when n=7, whereas for CV, the principal number has been limited to n=6. There are two reasons for this, one is the convergence as mentioned above. The other is the contribution from n=6 less than 0.04\%(CV). From table \ref{tab1}, we can see that the core-valence correlation is important in determining the energy of the calculated levels. The difference can be decreased to only 122 cm$^{-1}$. Our CV results agree better with experimental results than Hao's results, which only 3s$^2$3p and 3s3p$^2$ configurations were considered in the calculation. Energies for all 129 levels and splitting are listed in table \ref{tab:1}.

Dirac-Fock wave functions with a minimum number of radial functions are not sufficient to represent
the occupied orbitals. Extra configurations have to be added to adequately represent electron correlations
(i.e., mixing coefficients). These extra configurations are represented by CSFs and must have the same
angular momentum and parity as the occupied orbital\cite{Gillaspy2011}. For instance, the level 1s$^2$2s$^2$2p$^6$3s$^2$3p($^2P^-_{1/2})$ is represented by 0.9913 of 1s$^2$2s$^2$2p$^6$3s$^2$3p($^2P^-_{1/2})$ and 0.0890 of 1s$^2$2s$^2$2p$^6$3p$^3$($^2P^-_{1/2})$. The mixing coefficients for the
wave functions of some calculated levels are shown in Table \ref{tab2}. The most important contributions
to the total wave function of a given level are those from the same configuration. For example, the
configuration-mixed wave function for the 1s$^2$2s$^2$2p$^6$3s3p$^2$($^4P_{1/2}$ )level is represented as

 3s3p$^2$($^4P_{1/2}$)=0.96 3s3p$^2$($^4P_{1/2}$)+0.36 3s3p$^2$($^2S_{1/2})$

where 0.96 and  0.36 are the configuration mixing coefficients. Coefficients from less than 0.10
were calculated but are not explicitly given. Expansion coefficients for several levels from NIST \cite{NIST} are listed in Table \ref{tab2} for comparison. Also, the contribution from each level were listed in table \ref{tab2}. Take the 1s$^2$2s$^2$2p$^6$3s3p$^2$($^4P_{1/2}$ ) for example, 3s3p$^2$($^4P_{1/2}$)=0.83 3s3p$^2$($^4P_{1/2}$)+0.13 3s3p$^2$($^2S_{1/2})$, where 0.83, and 0.13 are contributions.
Clearly, the present and the previous \cite{NIST} results are
very close to one another in the description of the configuration-interaction wave functions. Because of the strong mixing, our original results for levels 9 and 11 were 3s3p$^2$($^2P_{3/2}$) and 3s$^2$3d($^2D_{3/2}$), which were different from experimental results. According to the NIST results, we adjusted the levels 9 and 11 to 3s$^2$3d($^2D_{3/2}$) and 3s3p$^2$($^2P_{3/2}$). For more complex systems it sometimes happens that two the same dominating LSJ term. The two levels will then get the same quantum labels in our output. The levels 15 and 31, levels 29 and 30, levels 38 and 40, levels 44 and 69, levels 48 and 60, levels 52 and 61, levels 56 and 79, levels 68 and 70, levels 72 and 73, levels 74 and 81, levels 88 and 92, levels 91 and 108, levels 93 and 106, levels 96 and 104 were the case, and the corresponding term were given by the main contribution.

A comparison between the present wavelengths and other published experimental results\cite{NIST} and theoretical results\cite{Huang1986,Hao2010} is shown in table \ref{tab3}. The accuracy of calculated CV wavelengths (in {\AA}) relative to measurements can be assessed from table \ref{tab3}, where the agreement is within 0.2{\AA}  for all available transitions except the transition 1-3 3s$^2$3p($^1P^o_{1/2}$)-3s3p$^2$($^4P_{1/2}$). 3s$^2$3p($^1P^o_{3/2}$)-3s3p$^2$($^2D_{3/2}$) with a calculated wavelength $\lambda$=140.76 {\AA}, which deviates from the measurement by about 0.01 {\AA}. The quoted experimental wavelength uncertainties are between 0.01{\AA}$\sim$ 0.03{\AA}. So the deviations actually reflect the estimated errors in the wavelengths. The difference between VV results and experimental results is the range of 0.08{\AA}$\sim$ 1.26{\AA}. Though a full set of transitions connecting to excited levels (3s$^2$3p) ground levels of Al-like Mo were performed by Huang \cite{Huang1986}, but we didn't use these data for comparison. Because of the level identification, Huang's results were different from ours and NIST. For example, wavelength for 3s$^2$3p($^1P^o_{1/2}$)-3s3p$^2$($^2S_{1/2}$) is 91.14 {\AA} by us, while this wavelength corresponded to the 3s$^2$3p($^1P^o_{1/2}$)-3s3p$^2$($^2P_{3/2}$) by Huang.

In appendix A,  many more Mo XXX E1
transitions in the soft X-ray region are listed than in any of the previous studies. The transition rate, the weighted oscillator strength and the line strength are given in Coulomb (velocity) and Babushkin (length) gauges. Also, for the electric
transitions the relative difference(dT)($dT=abs(A_l-A_v)/max(A_l,A_v)$) between the transition rates in length and
velocity gauges are given. A value close to dT=0 for an allowed
transition is a known accuracy indicator\cite{Ekman2014}. In many cases the
values are reasonably close to 0 but in other cases, for example the
15-81, 3p$^3$ $^2D_{1/2}$- 3s3d$^2$ ($^3P_2$)$^2P_{1/2}$
 transition, the difference can be larger than 0.9. In particular, our calculations presented in
appendix A provide comprehensive new data for M2 transitions for Mo XXX, which no existent data for public. This will help with
the identification of spectral lines of Mo XXX.

\section{Summary}
MCDHF and RCI calculations for 3s$^2$3p, 3s3p$^2$, 3s$^2$3d, 3p$^3$, 3s3p3d, 3p$^2$3d and 3s3d$^2$
configurations of Al-like Molybdenum are presented. Fine structure energy levels, oscillator strengths, line strengths, transition
probabilities and wavelengths for transitions among levels belonging to these levels are performed.
The valence-valence and core-valence correlation effects are accounted for in a systematic way.
The calculated energy levels and weighted oscillator strengths with core-valence correlation effect show a good agreement
with both theoretical and experimental data from the literature. The computed wavelengths are almost spectroscopic accuracy, aiding line identification in spectra. Uncertainties of the transition ares are estimated by dT, as suggested by Ekman \emph{et al}\cite{Ekman2014}. For most of the strong transitions, dT is below 0.1. For the weaker transitions, dT is somewhat larger, can up to 0.99. In addition, we have obtained some new
and previously unpublished energy levels for this ion. Our results are useful for many applications such
as controlled thermonuclear fusion, laser and plasma physics as well as astrophysics.

\ack

 This work was supported by the
National Natural Science Foundation of China (Grant No.11304266), Special Foundation for theoretical physics Research Program of China(Grant No.11547145) and Xuzhou Institute of technology(Grant No.XKY2015101).

\section*{Appendix A}
All the E1 and M2 transition data in the supplementary content. All the results are from CV calculations.
\section*{Appendix B Supplementarydata}
Supplementary data associated with this article can be found in the online version.

\clearpage
\begin{table}
\caption{Comparison between the present calculations of level energies (in cm$^{-1}$) and experimental data for some transitions in Mo XXX.$\Delta$1=VV-NIST, $\Delta$2=CV(n=6)-NIST , $\Delta$3=Hao-NIST.}
\label{tab1}
\begin{tabular}{llllllllllll}
\hline\noalign{\smallskip}
Level                & NIST     & CV(n=6) & CV(n=5)  & CV(n=4)  &  VV     &  DF     &    Hao    & $\Delta$1 & $\Delta$2 & $\Delta$3& $\Delta$4\\
\hline

$3s^23p$ $^2P^o_{1/2}$& 0       & 0       &  0       &   0      &  0      &       0 &     0       & 0    &  0    & 0      &  0 \\
$3s^23p$ $^2P^o_{3/2}$& 204020  & 204142  &  204135  & 204107   & 203856  &  207340 &    204106   & 122  & -164  &  3320  &  86  \\
$3s3p^2$ $^4P_{1/2}  $& 538435  & 539933  &  539777  & 539719   & 539383  &  540894 &    540083   & 1498 & 948   &  2459  &  1648\\
$3s3p^2$ $^2D_{3/2}  $& 816860  & 817241  &  817171  & 817397   & 818018  &  821549 &    819074   & 381  & 1158  &  4689  &  2214\\
$3s3p^2$ $^2P_{1/2}  $& 891280  & 891888  &  891997  & 892491   & 897138  &  895115 &    893363   & 608  & 5858  &  3835  &  2083\\
$3s3p^2$ $^2D_{5/2}  $& 914330  & 914569  &  914414  & 914409   & 913814  &  921564 &    916047   & 239  & -516  &  7234  &  1717\\
$3s^23d$ $^2D_{3/2}  $& 1080540 & 1081045 &  1081097 & 1081597  & 1085630 &  1087055&    $$       & 505  & 5090  &  6515  & $$    \\
$3s3p^2$ $^2S_{1/2}  $& 1095240 & 1097226 &  1097403 & 1098180  & 1100081 &  1103515&    1098614  & 1986 & 4841  &  8275  &  3374  \\
$3s3p^2$ $^2P_{3/2}  $& 1150820 & 1151578 &  1151755 & 1152664  & 1156119 &  1158211&    1117153  & 758  & 5299  &  7391  &  $$     \\
$3s^23d$ $^2D_{5/2}  $& 1162130 & 1162546 &  1162664 & 1163602  & 1164874 &  1170356&    $$       & 416  & 2744  &  8226  &  -44977\\

\noalign{\smallskip}\hline
\end{tabular}
\end{table}

\begin{table}
\caption{The configuration mixing coefficients and contributions for some level in Mo XXX. The number in the parenthesis refers to the level number(the key in Table \ref{tab:1}).}
\label{tab2}
\begin{tabular}{lllll}
\hline\noalign{\smallskip}
Key &   Level       &   Mix     &    Contribution    &    Reference\\
\hline
1 & 3s$^2$3p $^2P^o_{1/2}$  &   0.99(1) + 0.09(22) & 0.98(1) + 0.01(22)\\
2 & 3s$^2$3p $^2P^o_{3/2}$  &   0.99(2) + 0.10(30) & 0.98(2) + 0.01(30)\\
3 & 3s3p$^2$ $^4P_{1/2}$    &   0.91(3) + 0.37(10) & 0.83(3) + 0.14(10) & 0.82(3)+0.14(10)\\
4 & 3s3p$^2$ $^4P_{3/2}$    &  -0.99(4) + 0.16(8)  & 0.96(4) + 0.03(9)  \\
5 & 3s3p$^2$ $^4P_{5/2}$    &   0.80(5)-0.57(8)    & 0.64(5) + 0.33(8) \\
6 & 3s3p$^2$ $^2D_{3/2}$    &   0.86(6)+0.35(9)    & 0.74(6)+0.12(9)     & 0.76(6)+0.12(9)\\
7 & 3s3p$^2$ $^2P_{1/2}$    &   0.83(7)+0.43(10)   & 0.68(7) + 0.18(10) & 0.67(7)+0.21(10)\\
8 & 3s3p$^2$ $^2D_{5/2}$    &   0.72(8)+0.42(5)    & 0.52(8) + 0.34(5)  & 0.54(8)+0.34(5) \\
9 & 3s$^2$3d $^2D_{3/2}$    &   0.66(9)-0.73(11)   & 0.43(9)+053(11)  & 0.56(9)+0.43(11)\\
10& 3s3p$^2$ $^2S_{1/2}$    &   0.82(10)-0.51(7)   & 0.67(10) + 0.26(7) & 0.65(10)+0.30(7)\\
11& 3s3p$^2$ $^2P_{3/2}$    &   0.57(11)+0.65(9)   & 0.33(11)+0.42(9) & 0.32(11)+0.45(9)\\
12& 3s$^2$3d $^2D_{5/2}$    &   0.91(12)-0.37(8)   & 0.83(12)+ 0.14(8) & 0.86(12)+0.13(8)\\

\noalign{\smallskip}\hline
\end{tabular}
\end{table}

\begin{table}
\caption{Comparison between the present calculations of wavelengths ($\lambda$  in {\AA}) and experimental data for some transitions in Mo XXX. Diff1=CV-NIST, Diff2=VV-NIST.}
\label{tab3}
\begin{tabular}{llllllllll}
\hline\noalign{\smallskip}
i & j       &  NIST    &   CV & VV    &    Diff1  &    Diff2  \\
\hline
1&    3  & 186.22   & 185.21  & 185.40 & 1.01  &  0.82   \\
1&    6  & 122.42   & 122.36  & 122.25 & 0.06  &  0.17  \\
1&    7  & 112.17   & 112.12  & 111.47 & 0.05  &  0.70  \\
1&    9  & 92.55    & 92.5    & 92.11  & 0.05  &  0.44  \\
1&    10 & 91.27    & 91.14   & 90.90  & 0.13  &  0.37  \\
1&    11 & 86.86    & 86.84   & 86.50  & 0.02  &  0.36  \\
2&    6  & 163.17   & 163.11  & 162.82 & 0.06  &  0.35  \\
2&    7  & 145.5    & 145.54  & 144.24 & -0.04 &  1.26  \\
2&    8  & 140.77   & 140.76  & 140.85 & 0.01  &  -0.08 \\
2&    9  & 114.09   & 114.04  & 113.41 & 0.05  &  0.68  \\
2&    10 & 112.16   & 111.97  & 111.58 & 0.19  &  0.58  \\
2&    11 & 105.62   & 105.55  & 105.01 & 0.07  &  0.61  \\
2&    12 & 104.37   & 104.34  & 104.04 & 0.03  &  0.33  \\

\noalign{\smallskip}\hline

\end{tabular}
\end{table}

\clearpage
\datatables % This command is necessary to get the table names in toc
\LTright=0pt
\LTleft=0pt

\footnotesize
\begin{longtable}{@{\extracolsep\fill}llllll}
  \caption{MCDHF energy levels in Mo XXX .\label{tab:1}}
Key & configuration & Level &  J & Energies & Splitting\\
\hline
\endfirsthead
\multicolumn{6}{@{}c@{}}{Table \ref{tab:1} contd\dots}\\[6pt]
\hline \hline\noalign{\vskip3pt}
Key & configuration & Level & J & Levels & Splitting\\

% Nuclei&&&This Work&&This Work&&\\
%\noalign{\vskip3pt}
%Key & configuration & J & Levels & Splitting\\[5pt]
\hline
\endhead
\hline
\noalign{\vskip3pt}
\multicolumn{6}{r}{\itshape Continued\dots}\\
\endfoot
\noalign{\vskip3pt}\hline
\endlastfoot
  1  &  3s$^2$3p     & $^2P$           &   1/2 -  &        0.00  &          0.00  \\
  2  &  3s$^2$3p     & $^2P$           &   3/2 -  &   204142.17  &     204142.17  \\
  3  &  3s3p$^2$     & $^4P$           &   1/2 +  &   539932.66  &     335790.49  \\
  4  &  3s3p$^2$     & $^4P$           &   3/2 +  &   668580.17  &     128647.51  \\
  5  &  3s3p$^2$     & $^4P$           &   5/2 +  &   724214.07  &      55633.90  \\
  6  &  3s3p$^2$     & $^2D$           &   3/2 +  &   817241.14  &      93027.06  \\
  7  &  3s3p$^2$     & $^2P$           &   1/2 +  &   891888.45  &      74647.31  \\
  8  &  3s3p$^2$     & $^2D$           &   5/2 +  &   914569.39  &      22680.94  \\
  9  &  3s$^2$3d     & $^2D$           &   3/2 +  &  1081045.27  &     166475.88  \\
 10  &  3s3p$^2$     & $^2S$           &   1/2 +  &  1097225.55  &      16180.28  \\
 11  &  3s3p$^2$     & $^2P$           &   3/2 +  &  1151577.74  &      54352.19  \\
 12  &  3s$^2$3d     & $^2D$           &   5/2 +  &  1162545.98  &      10968.24  \\
 13  &  3p$^3$       & $^2D$           &   3/2 -  &  1403530.96  &     240984.98  \\
 14  &  3s3p3d       & ($^3P$)$^4F$    &   3/2 -  &  1501039.42  &      97508.46  \\
 15  &  3p$^3$       & $^2D$           &   5/2 -  &  1511880.34  &      10840.91  \\
 16  &  3p$^3$       & $^3S$           &   3/2 -  &  1549915.54  &      38035.20  \\
 17  &  3s3p3d       & ($^3P$)$^4F$    &   5/2 -  &  1557723.54  &       7808.00  \\
 18  &  3s3p3d       & ($^3P$)$^4F$    &   7/2 -  &  1606542.86  &      48819.33  \\
 19  &  3s3p3d       & ($^3P$)$^4D$    &   3/2 -  &  1628013.92  &      21471.06  \\
 20  &  3s3p3d       & ($^3P$)$^4P$    &   5/2 -  &  1630601.06  &       2587.14  \\
 21  &  3s3p3d       & ($^3P$)$^4D$    &   1/2 -  &  1630698.32  &         97.26  \\
 22  &  3p$^3$       &  $^2P$          &   1/2 -  &  1669684.43  &      38986.11  \\
 23  &  3s3p3d       & ($^3P$)$^2F$    &   5/2 -  &  1731519.59  &      61835.16  \\
 24  &  3s3p3d       & ($^3P$)$^4F$    &   9/2 -  &  1734429.20  &       2909.61  \\
 25  &  3s3p3d       & ($^3P$)$^2D$    &   3/2 -  &  1755461.64  &      21032.44  \\
 26  &  3s3p3d       & ($^3P$)$^4D$    &   5/2 -  &  1774961.58  &      19499.93  \\
 27  &  3s3p3d       & ($^3P$)$^4D$    &   7/2 -  &  1776545.64  &       1584.07  \\
 28  &  3s3p3d       & ($^3P$)$^4P$    &   1/2 -  &  1778672.85  &       2127.21  \\
 29  &  3s3p3d       & ($^3P$)$^4P$    &   3/2 -  &  1781174.25  &       2501.40  \\
 30  &  3p$^3$       &  $^2P$          &   3/2 -  &  1789984.63  &       8810.38  \\
 31  &  3s3p3d       & ($^3P$)$^2D$    &   5/2 -  &  1837623.51  &      47638.87  \\
 32  &  3s3p3d       & ($^1P$)$^2D$    &   3/2 -  &  1902843.56  &      65220.05  \\
 33  &  3s3p3d       & ($^3P$)$^2F$    &   7/2 -  &  1908301.16  &       5457.60  \\
 34  &  3s3p3d       & ($^1P$)$^2F$    &   7/2 -  &  1984804.08  &      76502.92  \\
 35  &  3s3p3d       & ($^1P$)$^2F$    &   5/2 -  &  1987248.47  &       2444.39  \\
 36  &  3s3p3d       & ($^3P$)$^2P$    &   1/2 -  &  1991520.71  &       4272.24  \\
 37  &  3s3p3d       & ($^1P$)$^2P$    &   1/2 -  &  2024759.51  &      33238.81  \\
 38  &  3s3p3d       & ($^3P$)$^2P$    &   3/2 -  &  2050128.33  &      25368.81  \\
 39  &  3s3p3d       & ($^1P$)$^2D$    &   5/2 -  &  2058040.17  &       7911.85  \\
 40  &  3s3p3d       & ($^1P$)$^2P$    &   3/2 -  &  2072367.02  &      14326.85  \\
 41  &  3p$^2$3d     & ($^3P$)$^4F$    &   3/2 +  &  2215298.66  &     142931.64  \\
 42  &  3p$^2$3d     & ($^3P$)$^4F$    &   5/2 +  &  2280269.86  &      64971.19  \\
 43  &  3p$^2$3d     & ($^1D$)$^2F$    &   5/2 +  &  2332185.03  &      51915.17  \\
 44  &  3p$^2$3d     & ($^3P$)$^2P$    &   3/2 +  &  2366071.26  &      33886.23  \\
 45  &  3p$^2$3d     & ($^1D$)$^2F$    &   7/2 +  &  2368026.03  &       1954.77  \\
 46  &  3p$^2$3d     & ($^3P$)$^4D$    &   1/2 +  &  2378107.68  &      10081.65  \\
 47  &  3p$^2$3d     & ($^3P$)$^4F$    &   7/2 +  &  2399202.37  &      21094.69  \\
 48  &  3p$^2$3d     & ($^3P$)$^4F$    &   9/2 +  &  2432360.86  &      33158.49  \\
 49  &  3p$^2$3d     & ($^3P$)$^4D$    &   3/2 +  &  2438520.74  &       6159.87  \\
 50  &  3p$^2$3d     & ($^3P$)$^4D$    &   5/2 +  &  2441684.31  &       3163.58  \\
 51  &  3p$^2$3d     & ($^1D$)$^2D$    &   5/2 +  &  2479568.11  &      37883.80  \\
 52  &  3p$^2$3d     & ($^3P$)$^4P$    &   3/2 +  &  2483257.49  &       3689.38  \\
 53  &  3p$^2$3d     & ($^3P$)$^4P$    &   1/2 +  &  2494586.01  &      11328.52  \\
 54  &  3p$^2$3d     & ($^1D$)$^2G$    &   7/2 +  &  2496986.85  &       2400.84  \\
 55  &  3p$^2$3d     & ($^3P$)$^2P$    &   1/2 +  &  2570478.79  &      73491.94  \\
 56  &  3p$^2$3d     & ($^1D$)$^2D$    &   3/2 +  &  2577258.59  &       6779.80  \\
 57  &  3p$^2$3d     & ($^3P$)$^2F$    &   5/2 +  &  2582017.35  &       4758.75  \\
 58  &  3p$^2$3d     & ($^3P$)$^4D$    &   7/2 +  &  2584888.65  &       2871.31  \\
 59  &  3p$^2$3d     & ($^1D$)$^2P$    &   1/2 +  &  2601588.98  &      16700.33  \\
 60  &  3p$^2$3d     & ($^1D$)$^2G$    &   9/2 +  &  2612466.36  &      10877.38  \\
 61  &  3p$^2$3d     & ($^3P$)$^4P$    &   3/2 +  &  2619122.62  &       6656.26  \\
 62  &  3p$^2$3d     & ($^3P$)$^4P$    &   5/2 +  &  2623100.42  &       3977.80  \\
 63  &  3s3d$^2$     & ($^3F_2$)$^4F$  &   3/2 +  &  2643749.20  &      20648.78  \\
 64  &  3s3d$^2$     & ($^3F_2$)$^4F$  &   5/2 +  &  2657868.79  &      14119.59  \\
 65  &  3s3d$^2$     & ($^3F_2$)$^4F$  &   7/2 +  &  2675504.83  &      17636.04  \\
 66  &  3s3d$^2$     & ($^3F_2$)$^4F$  &   9/2 +  &  2699614.93  &      24110.10  \\
 67  &  3p$^2$3d     & ($^3P$)$^2F$    &   7/2 +  &  2707035.71  &       7420.78  \\
 68  &  3p$^2$3d     & ($^1D$)$^2S$    &   1/2 +  &  2710674.63  &       3638.93  \\
 69  &  3p$^2$3d     & ($^1S$)$^2D$    &   3/2 +  &  2715983.07  &       5308.44  \\
 70  &  3s3d$^2$     & ($^3P_2$)$^4P$  &   1/2 +  &  2733451.18  &      17468.10  \\
 71  &  3s3d$^2$     & ($^1D_2$)$^2D$  &   3/2 +  &  2737356.61  &       3905.43  \\
 72  &  3s3d$^2$     & ($^3F_2$)$^4P$  &   5/2 +  &  2739241.99  &       1885.38  \\
 73  &  3p3d$^2$     & ($^1S_0$)$^2D$  &   5/2 +  &  2762209.41  &      22967.42  \\
 74  &  3p3d$^2$     & ($^3P_2$)$^2D$  &   5/2 +  &  2811197.77  &      48988.36  \\
 75  &  3s3d$^2$     & ($^1D_2$)$^2D$  &   3/2 +  &  2811626.31  &        428.54  \\
 76  &  3s3d$^2$     & ($^1G_2$)$^2G$  &   7/2 +  &  2836862.79  &      25236.48  \\
 77  &  3s3d$^2$     & ($^1G_2$)$^2G$  &   9/2 +  &  2840080.04  &       3217.25  \\
 78  &  3s3d$^2$     & ($^1D_2$)$^2D$  &   5/2 +  &  2842413.86  &       2333.83  \\
 79  &  3p3d$^2$     & ($^3P_2$)$^2D$  &   3/2 +  &  2856626.50  &      14212.64  \\
 80  &  3s3d$^2$     & ($^3F_2$)$^2F$  &   7/2 +  &  2925924.33  &      69297.83  \\
 81  &  3s3d$^2$     & ($^3F_2$)$^4F$  &   5/2 +  &  2927945.08  &       2020.75  \\
 82  &  3s3d$^2$     & ($^3P_2$)$^2P$  &   1/2 +  &  2951909.75  &      23964.68  \\
 83  &  3s3d$^2$     & ($^3P_2$)$^2P$  &   3/2 +  &  2996315.16  &      44405.41  \\
 84  &  3s3d$^2$     & ($^1S_0$)$^2S$  &   1/2 +  &  2998974.04  &       2658.88  \\
 85  &  3p3d$^2$     & ($^3F_2$)$^4G$  &   5/2 -  &  3156841.36  &     157867.31  \\
 86  &  3p3d$^2$     & ($^3F_2$)$^4G$  &   7/2 -  &  3215416.39  &      58575.04  \\
 87  &  3p3d$^2$     & ($^1D_2$)$^2F$  &   5/2 -  &  3249621.10  &      34204.71  \\
 88  &  3p3d$^2$     & ($^3F_2$)$^2D$  &   3/2 -  &  3266198.70  &      16577.60  \\
 89  &  3p3d$^2$     & ($^3F_2$)$^4G$  &   9/2 -  &  3272117.64  &       5918.93  \\
 90  &  3p3d$^2$     & ($^3P_2$)$^4D$  &   1/2 -  &  3288921.23  &      16803.59  \\
 91  &  3p3d$^2$     & ($^3P_2$)$^2S$  &   1/2 -  &  3291512.95  &       2591.72  \\
 92  &  3p3d$^2$     & ($^3F_2$)$^4F$  &   3/2 -  &  3310938.56  &      19425.61  \\
 93  &  3p3d$^2$     & ($^3F_2$)$^4D$  &   7/2 -  &  3330658.37  &      19719.81  \\
 94  &  3p3d$^2$     & ($^3F_2$)$^4F$  &   5/2 -  &  3347039.19  &      16380.82  \\
 95  &  3p3d$^2$     & ($^1G_2$)$^2H$  &   9/2 -  &  3350527.44  &       3488.25  \\
 96  &  3p3d$^2$     & ($^3P_2$)$^4D$  &   5/2 -  &  3359114.93  &       8587.49  \\
 97  &  3p3d$^2$     & ($^3P_2$)$^2D$  &   3/2 -  &  3373848.79  &      14733.86  \\
 98  &  3p3d$^2$     & ($^1G_2$)$^2G$  &   7/2 -  &  3375374.81  &       1526.02  \\
 99  &  3p3d$^2$     & ($^3P_2$)$^4P$  &   3/2 -  &  3387697.95  &      12323.14  \\
100  &  3p3d$^2$     & ($^3F_2$)$^4G$  &   1/2 -  &  3414888.95  &      27191.00  \\
101  &  3p3d$^2$     & ($^3F_2$)$^4F$  &   9/2 -  &  3452833.78  &      37944.83  \\
102  &  3p3d$^2$     & ($^3F_2$)$^4F$  &   7/2 -  &  3454687.42  &       1853.64  \\
103  &  3p3d$^2$     & ($^1D_2$)$^2P$  &   3/2 -  &  3464019.04  &       9331.61  \\
104  &  3p3d$^2$     & ($^3F_2$)$^2D$  &   5/2 -  &  3475632.89  &      11613.85  \\
105  &  3p3d$^2$     & ($^3F_2$)$^4D$  &   1/2 -  &  3482045.42  &       6412.54  \\
106  &  3p3d$^2$     & ($^3F_2$)$^4D$  &   7/2 -  &  3484382.56  &       2337.13  \\
107  &  3p3d$^2$     & ($^3F_2$)$^4D$  &   5/2 -  &  3484758.08  &        375.52  \\
108  &  3p3d$^2$     & ($^3P_2$)$^4P$  &   1/2 -  &  3499472.09  &      14714.01  \\
109  &  3p3d$^2$     & ($^3P_2$)$^4P$  &   7/2 -  &  3511956.89  &      12484.80  \\
110  &  3p3d$^2$     & ($^3F_2$)$^4D$  &   3/2 -  &  3512976.76  &       1019.87  \\
111  &  3p3d$^2$     & ($^1S_0$)$^2P$  &   1/2 -  &  3514651.06  &       1674.30  \\
112  &  3p3d$^2$     & ($^3P_2$)$^4P$  &   5/2 -  &  3523485.34  &       8834.28  \\
113  &  3p3d$^2$     & ($^1G_2$)$^2G$  &   9/2 -  &  3535403.24  &      11917.90  \\
114  &  3p3d$^2$     & ($^3F_2$)$^2F$  &   5/2 -  &  3552844.93  &      17441.68  \\
115  &  3p3d$^2$     & ($^1G_2$)$^2H$  &   1/2 -  &  3555979.73  &       3134.80  \\
116  &  3p3d$^2$     & ($^3P_2$)$^4S$  &   3/2 -  &  3560665.80  &       4686.07  \\
117  &  3p3d$^2$     & ($^1G_2$)$^2F$  &   7/2 -  &  3588231.79  &      27565.99  \\
118  &  3p3d$^2$     & ($^1D_2$)$^2D$  &   3/2 -  &  3610647.50  &      22415.71  \\
119  &  3p3d$^2$     & ($^1D_2$)$^2P$  &   1/2 -  &  3611504.92  &        857.42  \\
120  &  3p3d$^2$     & ($^1G_2$)$^2F$  &   7/2 -  &  3620870.30  &       9365.38  \\
121  &  3p3d$^2$     & ($^1D_2$)$^2D$  &   5/2 -  &  3643957.29  &      23086.99  \\
122  &  3p3d$^2$     & ($^3F_2$)$^2G$  &   9/2 -  &  3665893.11  &      21935.82  \\
123  &  3p3d$^2$     & ($^1G_2$)$^2F$  &   5/2 -  &  3686009.11  &      20116.00  \\
124  &  3p3d$^2$     & ($^1S_0$)$^2P$  &   3/2 -  &  3702659.72  &      16650.61  \\
125  &  3p3d$^2$     & ($^3F_2$)$^2G$  &   7/2 -  &  3713471.41  &      10811.69  \\
126  &  3p3d$^2$     & ($^3D_2$)$^2D$  &   3/2 -  &  3726988.84  &      13517.43  \\
127  &  3p3d$^2$     & ($^3P_2$)$^2D$  &   5/2 -  &  3742719.51  &      15730.67  \\
128  &  3p3d$^2$     & ($^3P_2$)$^2P$  &   1/2 -  &  3771330.94  &      28611.43  \\
129  &  3p3d$^2$     & ($^3P_2$)$^2P$  &   3/2 -  &  3809013.54  &      37682.60  \\

\end{longtable}

\end{document}